\documentclass[conference]{IEEEtran}
\IEEEoverridecommandlockouts

\usepackage{cite}
\usepackage{graphicx}
\usepackage{amssymb, amsmath}
\usepackage{url}

\begin{document}

\title{Uptime-Optimized Cloud Architecture as a Brokered Service}

\author{\IEEEauthorblockN{Sreekrishnan Venkateswaran}
\IEEEauthorblockA{IBM Corporation, India\\
Email: s\_krishna@in.ibm.com}
\and
\IEEEauthorblockN{Santonu Sarkar}
\IEEEauthorblockA{BITS Pilani, K K Birla Goa Campus, India\\
Email: santonus@goa.bits-pilani.ac.in}\thanks{S. Sarkar has been partially supported by a research grant from Accenture Technology Labs, USA.} }

\maketitle

\begin{abstract}
Enterprise workloads usually call for an uptime service level agreement (SLA) at the pain of contractual penalty in the event of slippage. Often, the strategy is to introduce ad-hoc HA (High Availability) mechanisms in response. Implemented solutions that we surveyed do not mathematically map their availability model to the required uptime SLA and to any expected penalty payout. In most client cases that we observed, this either resulted in an over-engineered solution that had more redundancies than was required, or in an inadequate solution that could potentially slip on the system uptime SLA stipulated in the contract. In this paper, we propose a framework backed by a model, to automatically determine the HA-enabled solution with the least TCO (total cost of ownership) for a given uptime SLA and slippage penalty. We attempt to establish that our work is best implemented as a brokered service that recommends an uptime-optimized cloud architecture.

\end{abstract}

\IEEEpeerreviewmaketitle

\section{Introduction}
Cloud Computing\footnote{The NIST definition of cloud
computing. Special Publication 800-145, 2011}, specifically Infrastructure as a service (IaaS), is a software-driven, algorithmic paradigm that supplies on-demand, pay-per-use compute capability with an upfront uptime assurance. 

There have been many efforts such as~\cite{lyu} from the academia and from the practice over several decades to design high availability on computer systems that propose new or improved methods for failover clustering. Papers such as \cite{Bassek} propose redundancy models for high availability of computer clusters. The work by \cite{Stanik} describes a clustered architecture that increases availability of middleware components. \cite{Shao} and \cite{Yerravalli} pertain to improving cluster failover mechanisms. White papers such as \cite{RedHat} describe clustering strategies specific to certain operating systems. 

However, to our knowledge, given a base cloud-hosted deployment architecture, there is no model-backed framework that has been proposed, which can automatically simulate and evaluate all HA (High Availability)-enabled variants of that base architecture, and determine the option that is optimal in terms of cost of clustering, given an uptime target and an associated financial burden to compensate non-adherence. 

Cloud environments are increasingly turning hybrid, and cloud service brokers further simplify IT consumption while also bringing flexibility of choice. In this paper we propose a framework backed by a probabilistic model that can be implemented by a cloud service broker to automatically provide an uptime-optimized cloud solution architecture as a service. A broker sits at a cross-cloud cross-customer vantage point, that we show, helps model and evaluate permutations of cluster technologies that constitute cloud-hosted systems. 

The paper has been organized as follows:
In Section~\ref{sec:modeling}, we first built a probabilistic model to construct expected uptimes for cloud-based systems that we consider as a serial combination of clustered compute, storage and network entities. We next simulate different levels of redundancies into these clusters and evaluate them based on the associated TCO (total cost of ownership). The TCO includes the cost of the infrastructure needed to embed redundancies, the labor needed to sustain the HA, and any penalty to be paid by the provider to compensate SLA slippage. Our proposed framework evaluates the TCO of all permutations of HA that can be engineered. The architecture that corresponds to the HA level that yields the least TCO is recommended as the optimal solution. 
We propose this framework to be realized as-a-service by a cloud broker since that is an entity at the best vantage point to gather values for the variables in our model. In Section~\ref{sec:evaluating}, we describe a client case study that establishes savings of up to 62\% by leveraging our technique. Section~\ref{sec:validity} examines any risks pertaining to the validity of our model evaluation. Finally, we conclude the paper and highlight future research directions.

\section {Modeling Brokered Uptime-aware Cloud Architecture}
\label{sec:modeling}
\begin{figure} 
\centering 
\includegraphics[scale=0.47]{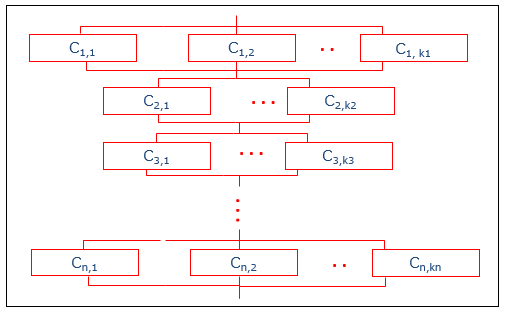}
\caption{Cloud-hosted Clustered IaaS Architecture of a System S}
\label{fig:cluster-model-on-cloud}
\vspace{-10pt}
\end{figure}

Let us model a cloud-hosted system $S$ as a serial combination of $n$ clusters Let each cluster $C_i$ be composed of $K_i$ nodes %each node denoted as $C_{i,ki}$ 
as shown in Figure~\ref{fig:cluster-model-on-cloud}.

\subsection{High Availability (HA) Solutions in Practice}
In a cloud based system, each cluster is built with the ``k-redundancy'' model.  %We use $C_i$ to denote the $i^{th}$ cluster in the system $S$ having $K_i$ nodes.  
Let $\widehat{K_i} \leq K_i$ be the maximum number of failed nodes that can be tolerated by the HA infrastructure of $C_i$. For instance, the case study shown later in this paper uses a VMWare ESX based HA solution with 3+1 cluster configuration, which means that the cluster has $K_i=4$ nodes, of which three nodes must be {\em active} for the cluster to be operational, with one standby node, i.e. $\widehat{K_i}=1$. If one of the operational nodes breaks down, the HA infrastructure, VMWare ESX in this case, brings the standby node to action. There is a small latency involved depending on the nature of the standby mode (hot, warm or cold) during which the cluster temporarily remains unavailable. This latency is commonly known as the {\em failover time}. The failover time $t_i$ (in minutes) for $C_i$ is a sum of (i) Time to detect that the currently active node in cluster $C_i$ is down (ii) Time to bring up the failover node if it is on standby and (iii) Time for the failover node to take over from the primary node.

Now consider the scenario when the number of node outages in $C_i$ exceeds $\widehat{K_i}$, which is 1 in this example. In that case, the cluster {\em breaks down} beyond immediate recovery. 

Because the cloud based system we consider is a serial combination of $n$ clusters, the entire system can {\em a)} Witness outage when one or more clusters break down completely or {\em b)} go down temporarily during a cluster failover.

\subsection{System Failure Model}
Since the above two events are mutually exclusive\footnote{Disregarding the possibility of an unrecoverable failure during cluster failover}, the overall downtime probability of $S$ can be expressed as: 
\begin{equation}
D_s = B_s +F_s
\label{eq:ds}
\end{equation}
where\\                
$B_s$ = System downtime probability due to breakdown of one or more clusters and\\
$F_s$ = System downtime probability due to failover latencies while clusters recover from node outages\\

\subsubsection{Calculating $\mathbf{B_s}$}
Let the probability that a node in cluster $C_i$ is down be $P_i$. %If the level of redundancy in a cluster is $N+\eta$, then $\widehat K_i$ is $\eta$. 
Then the probability that cluster $C_i$ is up is $\sum_{j=K_i-\widehat K_i}^{K_i} \binom{K_i}{j}(1-P_i)^{j}\cdot P_i^{K_i-j}$.

So system downtime probability 
\begin{equation}
B_s = 1-\prod\limits_{i=1}^{n}\sum\limits_{j=K_i-\widehat K_i}^{K_i} \binom {K_i}{j}(1-P_i)^{j}\cdot P_i^{K_i-j}
\label{eq:bs}
\end{equation}
\subsubsection{Calculating $\mathbf{F_s}$} Let $f_i$ be the average number of failures experienced by a node in cluster $C_i$ in a year. 
Total downtime due to failover transactions in cluster $C_i$ = $f_it_i(K_i-\widehat K_i)$ since there are $(K_i-\widehat K_i)$ active nodes in cluster $C_i$ at any given point of time.
Since $P_i$ is the probability that a node in cluster $C_i$ is down, it is also the probability that a {\em currently active node} in $C_i$ is down. However, multiple clusters might simultaneously be experiencing failover transactions. The downtime probability due to failover transactions in cluster $C_i$, when no other clusters are experiencing outages is $F_s(C_i)=\frac{f_it_i(K_i-\widehat K_i)}{\delta}P(X_i)$, where $X_i$ is the event that none of the clusters other than $C_i$ are experiencing failover events\footnote{We ignore the error of counting intra-cluster node failover times when more than $\widehat K_i$ nodes in $C_i$ fail simultaneously}, $P(X_i)$ is the probability of this event, and $\delta$ is 525600, the number of minutes in a year. Since $P(X_i)$ is the probability that all the currently running $(K_j-\widehat K_j)$ nodes in every cluster $C_{j, j \neq i}$ are up and running, $P(X_i) = \Pi_{j=1}^{n, j \neq i} [(1-P_j)^{K_j-\widehat K_j}]$. Substituting $P(X_i)$, we get $F_s(C_i)=\frac{f_it_i(K_i-\widehat K_i)}{\delta}\prod_{j=1}^{n, j \neq i}[(1-P_j)^{K_j-\widehat K_j}]$.
%Thus, downtime probability due to a failover transaction in cluster $C_i$ when there are no other simultaneous node outages in any other cluster is $
Generalizing, the system downtime probability due to all failover transactions across clusters 
\begin{equation}
F_s = \sum\limits_{i=1}^n F_s(C_i) = \sum\limits_{i=1}^n \frac{f_it_i(K_i-\widehat K_i)}{\delta}\prod_{j=1}^{n, j \neq i}[(1-P_j)^{K_j-\widehat K_j}]
\label{eq:fs}
\end{equation}
%\[F_s = \sum\limits_{i=1}^n (((f_i*t_i*(K_i-\widehat K_i))/\delta)*\prod\limits_{j=1}^{n, j \neq i}[(1-P_j)^{K_j-\widehat K_j}]) ~~~~~~~[3]
%\]

We can apply Equation~\ref{eq:bs} and~\ref{eq:fs} to compute the total downtime probability $D_s$ of system $S$ in Equation~\ref{eq:ds}.

Total uptime probability of $S$ is
\begin{equation}
U_s=1-D_s
\label{eq:us}
\end{equation} 
\subsection{Service Level Aware Model with Hybrid Cloud Brokerage}
\begin{figure} 
\centering 
\includegraphics[scale=0.30]{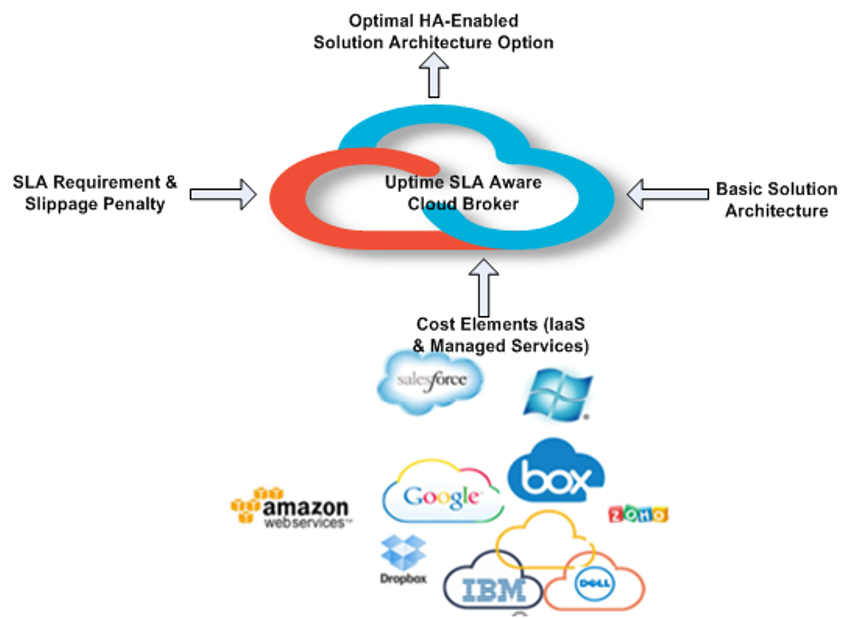}
\caption{Uptime Service Level Aware Cloud Brokerage}
\label{fig:brokered-uptime}
\vspace{-10pt}
\end{figure}

In the framework that we propose, a cloud service broker is provided the following inputs as shown in Figure~\ref{fig:brokered-uptime}:
\begin{enumerate}
\item The base cloud solution architecture in terms of the compute/storage/network topology on a specified underlying hybrid environment, mapped to customer requirements
\item The uptime SLA requirement $U_{SLA}$, expressed as a percentage, along with the contractual slippage penalty $SP$ per unit time (say per hour) of system unavailability
\item Being a hybrid cloud services broker, it is aware of the rate-carded price $C_{HA}$, which is the monthly infrastructure cost to engineer redundancy for the proposed level of HA plus the labor cost needed to deploy and sustain that HA for a period of one month
\end{enumerate}

By virtue of its vantage point above clouds and being the provisioning pane into these clouds across customers spanning a long timeline, the broker determines and maintains a database of  
\begin{enumerate}
\item The $P_i$ and $f_i$ across IaaS components across clouds
\item The $t_i$ for various components in the underpinning hybrid environment
\end{enumerate}

Using the above, the brokerage service models all permutations of HA-enabled system constructs ($k^n$ possibilities, where $n$ is the number of clusters and $k$ is the number of HA choices per cluster)  that conform to the input base solution architecture. Further, for each candidate system deployment $i$, it estimates the monthly $TCO_i$ (total cost of ownership) as the sum of (i) The cost to implement and sustain the proposed HA-enabled system and (ii) Any expected slippage penalty, which is calculated by determining expected system uptime using Equation~\ref{eq:us} and applying penalty clauses to projected downtime durations beyond the contractual SLA:
\begin{align}
TCO_i &= C_{HA} + (\frac{U_{SLA}}{100} - U_s)\frac{\delta}{12\times60}SP \nonumber\\
&= C_{HA} \mathrm{~when~} (\frac{U_{SLA}}{100} \leq U_s)\\
OptCh &= \min_{i=1,k^{n}}{TCO_i}
\end{align}

Note that $(\frac{U_{SLA}}{100} - U_s)\frac{\delta}{12\times60}$ converts downtime to slippage hours a month. The brokerage service then chooses the HA-enabled topology with the minimum TCO (i.e. $OptCh$) as the recommended HA strategy.

\section{Evaluating the Model with a Client Case Study}
\label{sec:evaluating}
\begin{figure} 
\centering 
\includegraphics[scale=0.29]{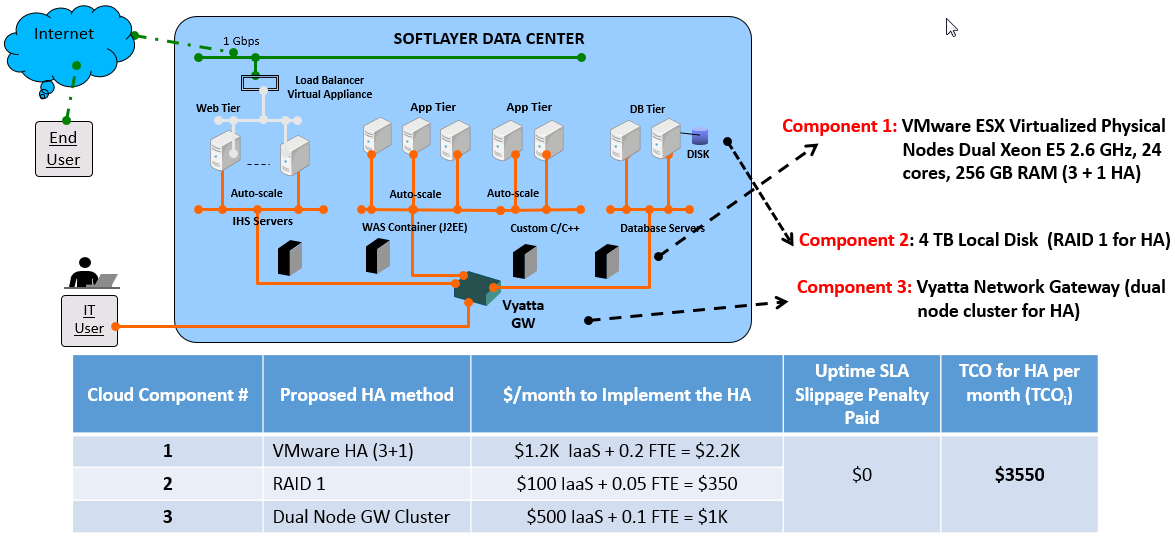}
\caption{HA with As-Is Strategy (Also Option \#8)}
\label{fig:as-is}
\vspace{-10pt}
\end{figure}

We considered a customer cloud solution designed as a three tier architecture comprising of application, database and web service tiers. The system, deployed on the IBM SoftLayer Cloud \cite{softlayer}, was a serial combination of 3 clusters at the compute, storage and network layers, respectively. The system uptime availability stipulated in the contract was 98\%; we assume a penalty of \$100 per hour of outage for illustration. In response, the provider added an adhoc level of HA in each of the three layers. The deployed solution and the cost to implement the HA is shown in Figure~\ref{fig:as-is}. The cost includes the monthly infrastructure cost of clustering on the SoftLayer cloud and the monthly labor (at \$30/hour in this case) to deploy and sustain the HA layers.

\subsection{Leveraging our Model}

\begin{figure} 
\centering 
\includegraphics[scale=0.28]{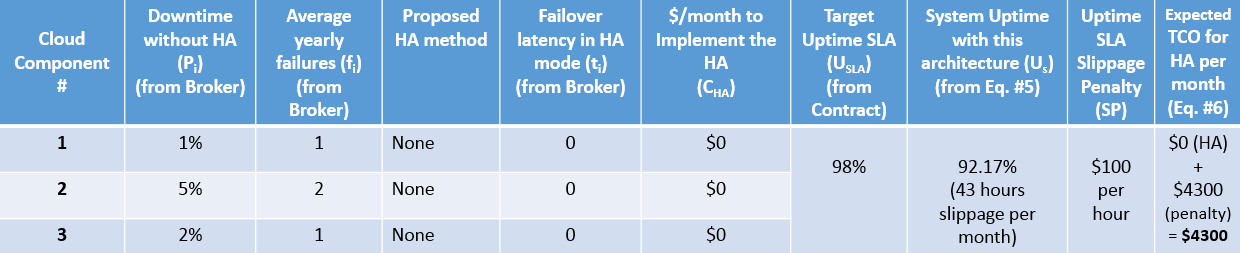}
\caption{Solution Option \#1: No HA in the System}
\label{fig:option-1}
\vspace{-10pt}
\end{figure}

\begin{figure} 
\centering 
\includegraphics[scale=0.28]{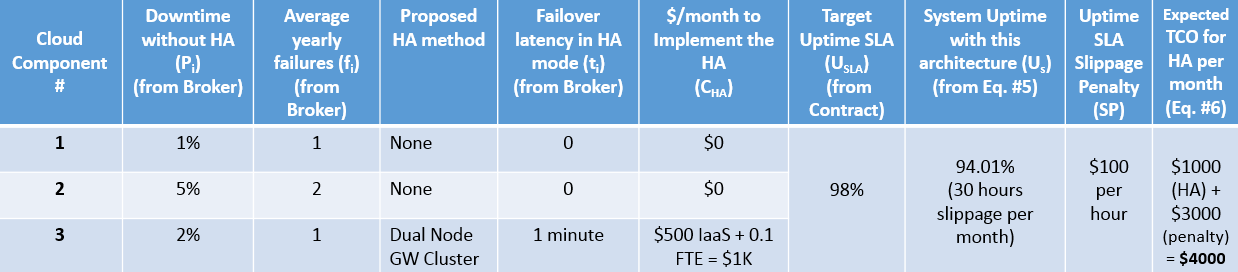}
\caption{Solution Option \#2: HA For Network Only}
\label{fig:option-2}
\vspace{-10pt}
\end{figure}

\begin{figure} 
\centering 
\includegraphics[scale=0.28]{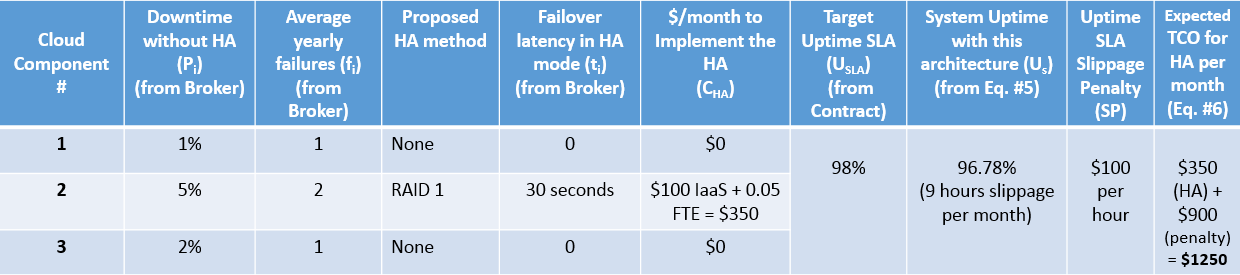}
\caption{Solution Option \#3: HA For Storage Only}
\label{fig:option-3}
\vspace{-10pt}
\end{figure}

\begin{figure} 
\centering 
\includegraphics[scale=0.28]{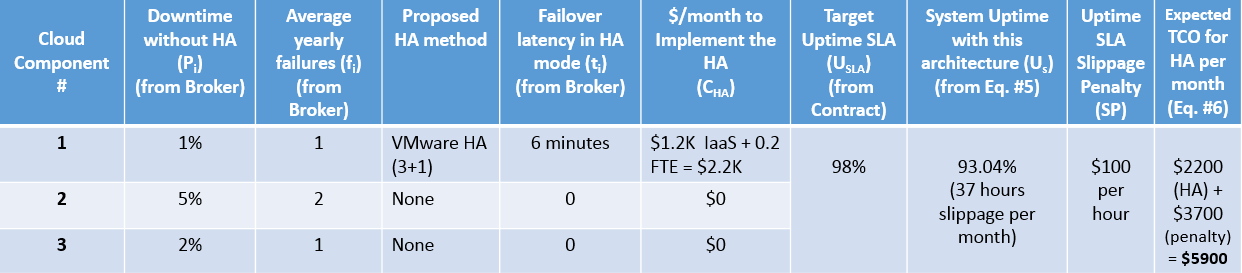}
\caption{Solution Option \#4: HA For Compute Only}
\label{fig:option-4}
\vspace{-10pt}
\end{figure}

\begin{figure} 
\centering 
\includegraphics[scale=0.28]{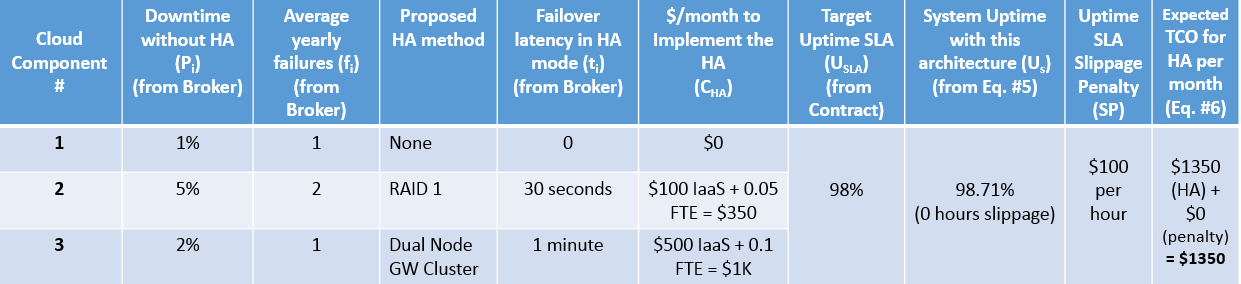}
\caption{Solution Option \#5: HA For Storage \& Network}
\label{fig:option-5}
\vspace{-10pt}
\end{figure}

\begin{figure} 
\centering 
\includegraphics[scale=0.28]{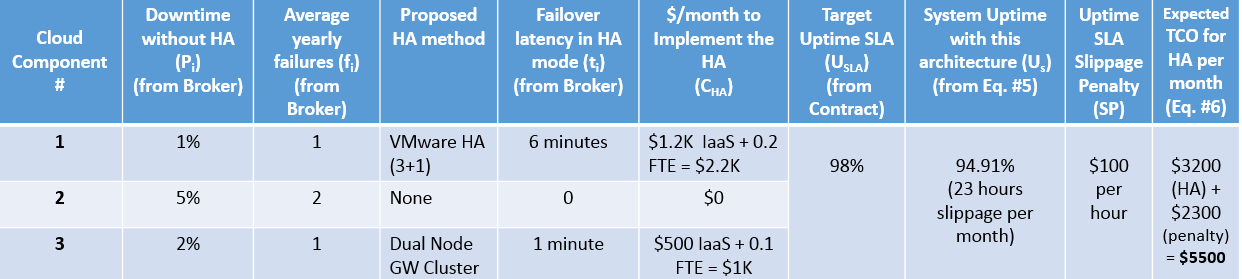}
\caption{Solution Option \#6: HA For Compute \& Network}
\label{fig:option-6}
\vspace{-10pt}
\end{figure}

We then applied our proposed framework that was implemented and embedded inside a cloud service broker. Our algorithm modeled all permutations of HA-enabled system constructs ($k^{n}$ possibilities, where, in our case $k=2$ and $n=3$) that conform to the input base 3-tier solution architecture. Figure~\ref{fig:option-1}, for instance, depicts a permutation where there is no HA or clustering in any component in the system; Figure~\ref{fig:option-2} is an embodiment of the same base architecture where only the network layer is clustered via dual gateways; the data in Figure~\ref{fig:option-3} is associated with an HA architecture where only the storage layer has redundancy via RAID-1 (Redundant Array of Independent Disks), and so on. For each candidate system instance, our algorithm, implemented as a brokered service, calculated the TCO. The service finally chose the HA-enabled topology with the minimum TCO as the recommended hosting strategy for the workload in question.
The $2^3$ solution options and their associated TCOs are show from Figure~\ref{fig:as-is} to Figure~\ref{fig:option-6} (Option \#7 that depicts HA for compute and storage is not shown other than in the tabular summary of Figure~\ref{fig:summary-of-results}). 

\subsection{Summary of Results \& Resulting Cost Efficiency} 
\begin{figure} 
\centering 
\includegraphics[scale=0.28]{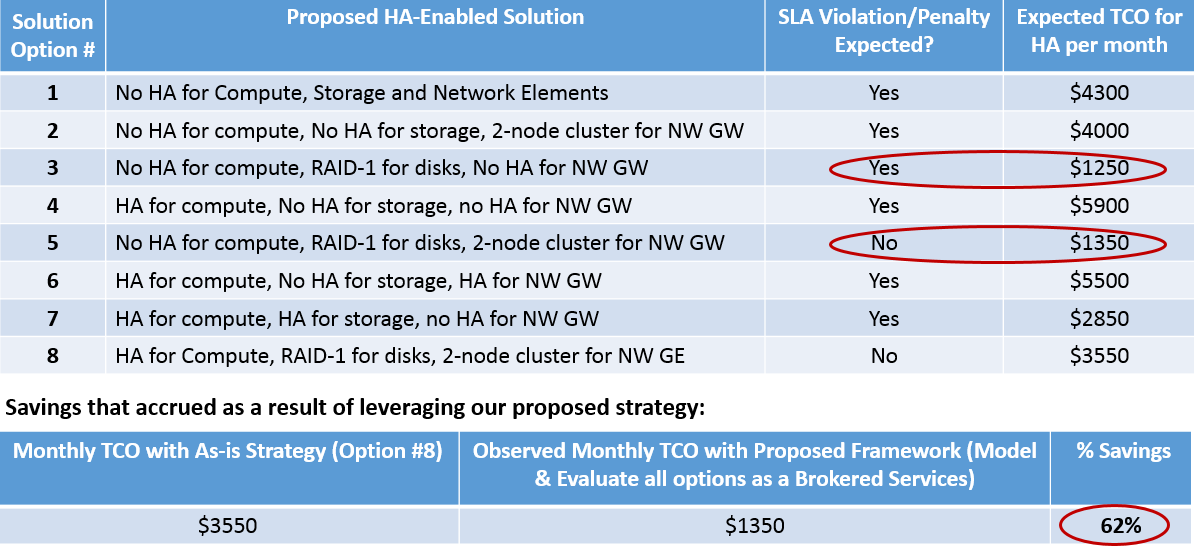}
\caption{Summary of Results \& Resulting Cost Efficiency}
\label{fig:summary-of-results}
\vspace{-10pt}
\end{figure}

In Figure~\ref{fig:summary-of-results}, we compare the TCOs between the as-is strategy (Figure~\ref{fig:as-is}) and our framework that simulated clustered scenarios, modeled associated TCOs, and determined the architecture with the minimum TCO. Our service thus evaluated all possible HA-enabled variants of the base configuration and applied them to Equation 6 to yield the recommended deployment, which is solution option \#3 (Figure~\ref{fig:option-3}). If the possibility of slippage penalty is to be minimized, the recommended option is \#5 (Figure~\ref{fig:option-5}). The savings that accrued as a result of applying our method was thus close to 62\%. Note that the actual savings realized will depend on how adhoc the original redundancy engineering has been.

\subsection{Improving the Time Complexity of the Algorithm} 
The technique we propose has exponential complexity. So if there are $n$ components in a cloud-hosted solution system that can potentially be clustered, the algorithm has a time complexity of $O(k^n)$. However, the value of $n$ in practice is usually low, most often under $10$ and the number of HA choices do not turn out to be significantly high.

The algorithm can be rendered more time efficient by pruning solution paths that are already optimized. The algorithm starts by evaluating all HA permutations where only one component is clustered, then proceeds to permutations where two components are clustered, and so on. If a particular permutation yields an uptime greater than what the contractual SLA stipulates, super-sets of that permutation can be pruned since those will increase uptime (beyond the SLA) while also increasing cost. For example, after evaluating option \#5, option \#8 can be clipped from the search tree without evaluation.

\section{Threats to Validity}
\label{sec:validity}
The following risks exist pertaining to \textit{construct validity}:
We rely on the cloud broker's long-term visibility across diverse deployments of multiple customers to gather and maintain values of $P_i$, $t_i$, and $f_i$. There could be skews in these values influenced by the demand and supply commerce pertaining to the broker's marketplace, but over the long-term, any contortions should smooth out. Also, the reliability of the broker has not been considered in our model.

On risks pertaining to \textit{internal validity}, SLA slippage penalty is a techno-commercial aspect, so there could be non-technical aspects at play when a provider fashions a strategy to address an SLA. This paper does not venture into territory outside of mathematics and technology.

We have attempted to establish \textit{external validity} by applying our model to one family of solutions that apply hypervisor-level clustering for compute, RAID for data stores, and dual clustered gateways for network. In future work, we plan to expand our research and further generalize our conclusions by applying to more customer case studies and broader HA strategies such as described in the next section. 

\section{Conclusion and Future Work}
In this paper, we proposed a framework and a supporting model to automatically simulate and evaluate all HA-enabled variants of a given base cloud architecture and determine the solution option that is minimalist in terms of total cost of ownership, given an uptime target. We also showed that cloud brokers are at good vantage point to leverage our framework and provide an uptime-optimized cloud solution architecture as a service. We applied our framework to a client case study and determined that our technique, had it been applied, would have resulted in a 62\% savings on the cost that was expended to implement high availability.

Our model and framework needs to be tested on complex scenarios that include OS clustering for compute, Software Defined Storage and clustered file systems for data stores, multi-pathing for storage I/O, and BGP (Border Gateway Protocol) over dual circuits for network redundancies. Moreover, though our technique is agnostic of the underpinning cloud and managed service environments, our testing has been on a single cloud (IBM SoftLayer) and not on a hybrid environment. We propose to address these in future work.
The larger goal of our research is to design what we envisage as next-generation cloud brokerage that constructs a commercial meta-cloud whose ownership is scattered across cloud providers.

\small{
\bibliographystyle{abbrv}
\bibliography{cloud.krish}
}

\end{document}